\documentstyle [12pt,a4p,epsfig,amsmath,multicol]{article}
\begin{document}
\vspace*{0.6cm}
\begin{center}
{\bf On the Einstein-Podolsky-Rosen Proof of the `Incompleteness'
 of Quantum Mechanics }
\end{center}
\vspace*{0.6cm}
\centerline{\footnotesize J.H.Field}
\baselineskip=13pt
\centerline{\footnotesize\it D\'{e}partement de Physique Nucl\'{e}aire et 
 Corpusculaire, Universit\'{e} de Gen\`{e}ve}
\baselineskip=12pt
\centerline{\footnotesize\it 24, quai Ernest-Ansermet CH-1211 Gen\`{e}ve 4. }
\centerline{\footnotesize E-mail: john.field@cern.ch}
\baselineskip=13pt
 
\vspace*{0.9cm}
\abstract{ It is shown that the Einstein-Podolsky-Rosen conclusion
 concerning the `incompleteness' of Quantum Mechanics does not
 follow from the results of their proposed gedanken experiment,
 but is rather stated as a premise. If it were possible to 
 perform the experiment it would, in fact, show that  Quantum 
 Mechanics is `complete' for the observables discussed.
 Because, however, of the non square-integrable nature of the 
 wave function, the proposed experiment gives vanishing
 probabilities for measurements performed in finite intervals 
 of configuration or momentum space. Hence no conclusion as
 to the `completeness', or otherwise, of Quantum Mechanics can
 be drawn from the experiment.}
\vspace*{0.9cm}
\normalsize\baselineskip=15pt
\setcounter{footnote}{0}
\renewcommand{\thefootnote}{\alph{footnote}}
\newline
 PACS 03.65.-w

\vspace*{0.4cm}

\newpage
\par Perhaps no other paper written in the present century has generated
as much debate about questions related to the foundations of physics and 
their philosophical implications than that of Einstein, Podolsky and
Rosen (EPR)~\cite{x1}. However, after the initial replies written by
Bohr~\cite{x2}, Furry~\cite{x3} and Schr\"{o}dinger~\cite{x4}, there
has been very little critical discussion of the EPR paper itself in
the literature~\cite{x5}. In this letter a reappraisal of the EPR paper
 is made and the following conclusions are drawn:
\begin{itemize}
\item[(i)] The argument presented by EPR to demonstrate the 
`incompleteness' of Quantum Mechanics (QM) is invalidated by a logical
 error.
\item[(ii)] The gedanken experiment proposed by EPR cannot be carried
 out if the usual probabilistic interpretation of QM is correct, and
 so no physical conclusions can be drawn from the experiment.
\end{itemize}
\par Following EPR, a theory is said to give a `complete' description
of a physical quantity if the following condition is satisfied:
\par{\it `Without, in any way, disturbing a system, we can predict with
 certainty (i.e. with probability equal to unity) the value of the 
 physical quantity'.}
\par If this is the case, EPR associate an `Element of Physical Reality'
to the the corresponding quantity. EPR also require that, in a `complete'
theory:
\par{\it `Every Element of Physical Reality must have a counterpart
 on the physical theory'.}
\par This hypothesis is not particularly important since it must
  necessarily be true if the theory is able to predict the value of
 the corresponding physical quantity. 
\par The EPR gedanken experiment will first be discussed from a 
purely logical viewpoint. Secondly, the conceptual feasiblity within
QM, of the proposed experiment is examined. The following hypotheses
 are defined:
\begin{itemize}
\item  $QMT$~: QM is a true theory within its domain of applicablity.
\item $QMTC(A,B,..)$~: QM is a true, complete, theory for the physical
 quantities A,B,.. .
\item $PRNC(A,B)$~: Elements of Physical Reality exist for each of a 
      pair of physical quantities A, B  with non-commuting operators
      in QM.
\end{itemize}
 The EPR gedanken experiment is based solely on the hypothesis QMT
 (Quantum Mechanics True). Contrary to the statement of EPR, it is
\underline{not necessary} to assume at the outset that QM is also a
complete theory (hypothesis QMTC). In fact, applying QMT and assuming
also that a quantum mechanical system of two correlated particles with
a certain well-defined wave function can be constructed, EPR found 
that Elements of Physical Reality 
 apparently \underline{can} be assigned to each 
of the quantities $P$ and $Q$ that that have non-commuting operators.
Using the symbol $\Rightarrow$ for `logically implies' EPR then found
that:
\[ QMT~\Rightarrow PRNC(P,Q) \]
 After correction~\cite{x6}, the final statement of the result of the
 gedanken experiment is:
 \par{\it `Starting from the assumption of the correctness of QM
 (i.e. hypothesis QMT) we arrived at the conclusion that two physical
  quantities with non-commuting observables can have simultaneous
  reality.'}
\par According to EPR's definitions,
if two physical quantities have corresponding elements of physical
 reality, then the theory is a complete one for these quantities.
 In symbols~\cite{x7}:
\begin{equation}
 PRNC(P,Q) \otimes QMTC(P,Q) = TRUE 
\end{equation}
Using De Morgan's Theorem, (1) implies:
\begin{equation}
\overline{PRNC(P,Q)} \oplus \overline{QMTC(P,Q)} = FALSE 
\end{equation}
However, EPR state that the right side of (2) is TRUE and conclude,
 instead of (1), that:
\begin{equation}
 PRNC(P,Q) \otimes QMTC(P,Q) = FALSE 
\end{equation}
Since the gedanken experiment showed that:
\[ PRNC(P,Q)=TRUE, \]
 EPR drew, on the basis of (3), the erroneous conclusion that:
\[ QMTC(P,Q)=FALSE. \]
i.e. that QM is an incomplete theory. 
 The basic assertion of EPR (actually, as shown above, in contradiction
to the result of their gedanken experiment) is:
\begin{equation}
\overline{PRNC(P,Q)}\oplus\overline{QMTC(P,Q)} = TRUE 
\end{equation} 
How is this assertion justified in the EPR paper? After discussion of
quantum mechanical measurements on a {\it single particle}, with no
obvious relevance to the case of {\it two correlated particles}
as used in their gedanken experiment, EPR state that:
\par {\it `From this it follows [1] the quantum mechanical description
 of reality given by the wave function is not complete or [2] when the
 operators corresponding to two physical quantities do not commute the 
 two quantities cannot have simultaneous reality.  For if both of
 them had simultaneous reality - and thus definite values - these values
 would enter into the complete description according to the condition
 of completeness. If the wave function provided such a complete 
 description of reality it would contain these values, these would be 
 predictable. This not being the case we are left with the
 alternatives stated.'}
 \par This argument, expressed symbolically by Eqn.(4),
  seems to be justified
 by the preceding discussion in the paper of non commuting observables
 (position and momentum) for a {\it single } particle, not the correlated
 two particle system of the gedanken experiment subsequently presented.
 In fact no justification is given by EPR for the application,
 {\it a priori}, of propositions [1] and [2] to the gedanken experiment.
 Even so, one can still ask what is the meaning of EPR's assertion in
 Eqn.(4)?  As quoted above, EPR carefully explain that the proposition
 [2] implies that the quantum mechanical description of two 
 commuting observables is not complete, i.e.
\[  [2] \equiv \overline{PRNC(A,B)} \Rightarrow \overline{QMTC(A,B)} \]
 But proposition [1] {\it is} $\overline{QMTC(A,B)}$, so the assertion
 of EPR is actually `either quantum mechanics is not complete or 
 quantum mechanics not complete', so that their conclusion that 
quantum mechanics is not complete is inevitable!
 The assertion only become logically
 coherent in the case that `not complete' in proposition [1] is replaced
 by `complete' equivalent to the always correct assertion 
`either quantum mechanics is complete or 
 quantum mechanics not complete'.
 \par To summarise, the EPR argument is based on the logical proposition:
   \[ X \oplus Y = TRUE \]
  It follows from this that if $X = FALSE$ then $Y = TRUE$ or {\it vice versa}.
   It is also possible that $X$ and $Y$ are both true; indeed the only
    case exluded by the proposition is that $X$ and $Y$ are both false.
   In their interpretation of the proposition EPR exclude the 
   possiblity that  $X$ and $Y$ are both true, in which case the only remaining
    ones are $X = FALSE$ and $Y = TRUE$ or $Y = FALSE$ and $X = TRUE$.
    But the truth (or falsehood) of the proposition
      $X \equiv \overline{PRNC(P,Q)}$ entails the truth (or falsehood) of
        the proposition
      $Y \equiv \overline{QMTC(P,Q)}$. Therefore if $X$ is false --the 
     claimed conclusion of EPR's analysis of their gedanken experiment--
      $Y$ must also be false, so that QM is then a complete theory,
     not an incomplete one as claimed by EPR. In fact the actual conclusion,
    that both $X$ and $Y$ are false, is the only one that is always
     inconsistent with EPR´s initial proposition. As pointed out above, when 
     $X$ and $Y$ are both false the `$TRUE$' on the right side of EPR's initial
     proposition must be replaced by `$FALSE$'. 
 \par Correcting this logical error, it might seem that the EPR 
 experiment establishes the `completeness' of quantum mechanics for
 the two non-commuting quantities P and Q. For this, however, it is
necessary that the suggested gedanken experiment
  can, at least in principle,
 be performed. It will now be shown that this is not the case, so 
 that no conclusion can be drawn as the the `completeness', or
 otherwise, of quantum mechanics, by the arguments presented by
 EPR.
\par The spatial wave function of the correlated two particle system
discussed by EPR is:
\begin{equation}
\Psi(x_1,x_2) = \int_{-\infty}^{\infty}dp \exp {\frac{2 \pi i}{h}(x_1-x_2+x_0)p}
  = h \delta(x_1-x_2+x_0)
\end{equation}
 The probability that the particle 1 will be observed in the interval
 $a < x_1 <b$, for any position of the particle 2, can be written as: 
\begin{eqnarray}
 P(a < x_1 <b) & = &  Lim (L \rightarrow \infty)
\frac{\int_{a}^{b} dx_1 \int_{-\infty}^{\infty} dx_2
|\Psi(x_1,x_2)|^2}
{\int_{-L}^{L} dx_1 \int_{-\infty}^{\infty} dx_2
|\Psi(x_1,x_2)|^2} \nonumber \\
 & = &  Lim (L \rightarrow \infty) \frac{b-a}{2L} = 0     
\end{eqnarray}
 The particle 1 cannot therefore be observed in any finite
 interval of $x_1$, and so the $Q$ measurement suggested in the EPR
 gedanken experiment cannot be carried out.
\par By making Fourier transforms with respect to $x_1$ and $x_2$
the momentum wavefunction corresponding to (5) is found to be:
\begin{equation}
\Psi(p_1,p_2) = \frac{h^2}{2 \pi} \exp \frac{2 \pi i p_1 x_0}{h}
 \delta(p_1+p_2)
\end{equation}
The probability to observe $p_1$ in the range $p_a <  p_1 < p_b$ for
 any value of $p_2$ is:
\begin{eqnarray}
 P(p_a < x_1 < p_b) & = &  Lim (p \rightarrow \infty)
\frac{\int_{p_a}^{p_b} dp_1 \int_{-\infty}^{\infty} dp_2
|\Psi(p_1,p_2)|^2}
{\int_{-p}^{p} dp_1 \int_{-\infty}^{\infty} dp_2
|\Psi(p_1,p_2)|^2} \nonumber \\
 & = &  Lim (p \rightarrow \infty) \frac{p_b-p_a}{2p} = 0     
\end{eqnarray}
 The momentum of particle 1 cannot be measured in any finite interval
 so that the proposed $p_2=P=-p_1$ measurement of the EPR gedanken
 experiment cannot be carried out. In fact, the correlated two particle
 wave function proposed by EPR is not square integrable either in 
 configuration or momentum space and so has no probabilistic 
 interpretation in QM. The single particle wavefunction discussed by
 EPR has the same shortcoming. Hence the `relative probability'
 $P(a,b)$ of EPR's Equation (6) also vanishes. While the statement
 that `all values of the coordinate are equally probable' is true,
 it is also true that the absolute probablity to observe the
 particle in any finite interval is zero. 
\par The EPR two particle wavefunction is now modified to render it
 square integrable so that the results of the gedanken experiment may be
 interpreted according to the usual rules of QM. The suggested
`minimally modified' wavefunction is:
\begin{equation}
\tilde{\Psi}(x_1,x_2) = \frac{1}{(\sqrt{2\pi}\sigma_x)^{\frac{1}{2}}}
 \exp \left( {\frac{x_0^2-2x_1^2-2x_2^2}{16 \sigma_x^2}} \right)
 \delta(x_1-x_2+x_0)
\end{equation}
Like the EPR wavefunction (5) $\tilde{\Psi}$ vanishes unless $x_2=x_1+x_0$,
but it is square integrable and normalised:
\begin{equation}
\int_{-\infty}^{\infty}\int_{-\infty}^{\infty}|\tilde{\Psi}(x_1,x_2)|^2
dx_1dx_2 = 1
\end{equation}
The EPR wavefunction (5) is recovered in the limit $\sigma_x \rightarrow
\infty$. Performing a double Fourier transform on Eqn(9) yields the
 corresponding momentum wave function:
\begin{equation}
\tilde{\Psi}(p_1,p_2) = \frac{1}{\pi \sigma_p}
\exp \left( {-\frac{(p_1+p_2)^2}{2 \sigma_p^2}}\right)
 \exp \left(\frac{2 \pi i p_1 x_0}{h}\right) 
\end{equation}
 where
 \[ \sigma_p = h/4 \pi \sigma_x \].
 The wavefunction (11) is also square integrable and normalised:
\begin{equation}
\int_{-\infty}^{\infty}\int_{-\infty}^{\infty}|\tilde{\Psi}(p_1,p_2)|^2
dp_1dp_2 = 1
\end{equation}
and the EPR wavefunction (7) is recovered in the limit
 $\sigma_x \rightarrow \infty$,
 $\sigma_p \rightarrow 0$. Now, performing the EPR gedanken experiment,
using instead the wavefunctions (9) and (11) it becomes clear that it 
is no longer possible to associate `Elements of Physical Reality' to the
position Q and the momentum P of the second particle by performing
measurements on the first one. The probability $\delta P(x_1)$ that
the spatial position of the first particle lies in the interval
 $\delta x_1$ around $x_1$\cite{x8} is:
\begin{equation}
\delta P(x_1) = \frac{1}{\sqrt \pi \sigma_x} \exp \left(
-\frac{x_0^2}{8 \sigma_x^2}\right)
 \exp \left(-\frac{(2x_1+x_0)^2}{8 \sigma_x^2} \right) \delta x_1
\end{equation}
Because of the $\delta$-function in the wave function (9), this is
also the probability that $x_2$ lies in the interval of width
$\delta x_2 =\delta x_1$ around $x_2 = x_1 +x_0$. Measuring $x_1$ 
in the interval $\delta x_1$ then enables the certain prediction
that $x_2$ lies in the interval $\delta x_2$ around $x_2 = x_1 +x_0$.
However, to associate an `Element of Physical Reality' to $x_2$
requires that the
 {\it value} must be exactly predictable. For this it is necessary
that $\delta x_1 = \delta x_2 \rightarrow 0$. In this case $\delta P(x_1)$
vanishes and no possiblity exists to measure the position of the
particle 1. The situation is then the same as in the case of the 
original EPR wavefunction (5). It is then clear that the product of the
uncertainties in P and Q can be much smaller than that required by the
Heisenberg Uncertainty Principle. However in order to thus determine Q, 
use is made of of the precise knowledge of the parameter $x_0$ of
the wavefunction, i.e. exact knowledge of how the wavefunction is 
prepared is required. But if {\it a priori} knowledge about
wavefunction preparation is admitted, it is trivial to show that
 observables with non-commuting operators can be simultaneously
 `known' with a joint precision far exceeding that allowed by the
Heisenberg Uncertainty Relation. To give a concrete example of this,
the process of para-positronium annihilation at rest: $ e^+e^-
 \rightarrow \gamma \gamma$ may be considered. The uncertainty in
 the momentum 
$\Delta p$ of one of the decay photons is determined by the mean
 lifetime of the decay process $\tau = 1.25 \times 10^{-10}$ sec:
 \[\Delta p = \frac{h}{c \tau}. \]
 The Heisenberg Uncertainty Relation then predicts 
 \[ \Delta x > 3.75 cm. \]
 The technically simple measurement of the position of the photon
 in the direction parallel to its momentum to within
 1mm (for example, by observing a recoil electron from Compton
 Scattering of the photon~\cite{x9}) then allows simultaneous 
 knowledge of the position and momentum of the electron
 (whose quantum mechanical operators do not commute) with 
 an accuracy $\simeq$ 40 times better than `allowed' by the
 Heisenberg Uncertainty Relation. Of course the Uncertainty Relation
 does indeed limit the precision of any attempt to 
 {\it simultaneously measure} a pair of non-commuting observables.
 However, as the counter example given above shows, it does not
 apply to {\it a priori} knowledge from state preparation,
 as used by EPR in the discussion of their gedanken experiment.
 There is therefore nothing remarkable (certainly no `paradox')
 in the fact that non-commuting observables can be `known' more 
 accurately than allowed by the Uncertainty Relation if information
 about state preparation is also included, as is the case for the
 EPR gedanken experiment.In fact, information from state preparation
 is essential for the EPR analysis of (hypothetical) measurements
  of the system described by the wavefunction (5). According to the
  latter the value of $x_2$ is fixed by a measurement of $x_1$ and {\it vice versa}.
   In both cases the information on the unmeasured variable is given by
   prior knowledge of the prepared wavefunction of the system.  
\par It has been stressed above, that no meaningful conclusions can be 
drawn from any  gedanken experiment based upon non square-integrable
wave functions. A similar criticism was made by Johansen
~\cite{x10} concerning a paper of Bell~\cite{x11} where the erroneous
conclusion was drawn, by the use of a non square-integrable
wave function, that states with a positive Wigner distribution
 (as is in fact the case for the EPR wave function (5)) necessarily
yield a local hidden variable model. A corollary is given by the
`complementary' limits discussed by Bohr~\cite{x2}, where an aspect
 of classical physics is recovered, yielding a precise position or
 momentum for a particle. Such exact limits are of limited physical
interest since the corresponding wavefuctions are not square 
 integrable for the conjugate variable, and so can have no
 physical interpretation within quantum mechanics. The Dirac
 $\delta$-function is a calculational device of extreme utility.
 It should never be forgotten, however, that it is only a mathematical
 idealisation never realised in the wavefunction of any actual
 physical system. 
\newline
{\bf Acknowledgements}
\par I thank N.Gisin and D.J.Moore for reading this paper and for
 their critical comments. 

\pagebreak

\pagebreak 

\begin{thebibliography}{99}
\bibitem{x7}
Each hypothesis is assumed to be either true or false. The
symbols $\otimes$ and $\oplus$ denote, respectively logical `and'
 and  `or'. A bar on a logical variable indicates negation.
\end{thebibliography}
\end{document}